\documentclass[prc,preprint,showpacs,showkeys,nofootinbib,tightenlines]{revtex4}
\usepackage{graphicx}  %
\usepackage{bm}  %                                          

\newcommand{\beq}{\begin{equation}}
\newcommand{\eeq}{\end{equation}}
\newcommand{\bea}{\begin{eqnarray}}
\newcommand{\eea}{\end{eqnarray}}
\newcommand{\ben}{\begin{eqnarray*}}
\newcommand{\een}{\end{eqnarray*}}

\newcommand{\simle}{\hspace*{0.2em}\raisebox{0.5ex}{$<$}
     \hspace{-0.8em}\raisebox{-0.3em}{$\sim$}\hspace*{0.2em}}

\renewcommand{\vec}[1]{{\mathbf #1}} 
\def\vecsigma{\mbox{\boldmath $\sigma$}}

\newcommand{\nab}{\overrightarrow{\nabla}}

\newcommand{\galnab}{\tensor{\nabla}}

\newcommand{\Mlo}{M_{lo}}
\newcommand{\Mhi}{M_{hi}}

\begin{document}

\title{Effective Field Theory for Halo Nuclei: Shallow $p$-Wave States}

\author{C.A. Bertulani}%\email{bertulani@nscl.msu.edu}
\affiliation{National Superconducting Cyclotron Laboratory,
             Michigan State University, East Lansing, MI\ 48824, USA}
\author{H.-W. Hammer}%\email{hammer@mps.ohio-state.edu}
\affiliation{Department of Physics,
         The Ohio State University, Columbus, OH\ 43210, USA}
\author{U. van Kolck}%\email{vankolck@physics.arizona.edu}
\affiliation{Department of Physics, University of Arizona,
          Tucson, AZ\ 85721, USA}
\affiliation{RIKEN-BNL Research Center, Brookhaven National Laboratory,
           Upton, NY\ 11973, USA}

%\today
\date{August 8, 2002}

\begin{abstract}
Halo nuclei are a promising new arena for studies based on effective 
field theory (EFT).
We develop an EFT for shallow $p$-wave states 
and discuss the application to elastic $n\alpha$ scattering.
In contrast to the $s$-wave case, both the scattering length and
effective range enter at leading order.
We also discuss the prospects of using EFT in
the description of other halos, such as the three-body
halo nucleus $^6$He.
\end{abstract}

\smallskip
\maketitle  

\section{Introduction}
\label{sec:intro}

Nuclear halo states have been found in a number of light nuclei 
close to the nucleon drip lines. They are characterized by a 
very low separation energy of the valence nucleon (or cluster 
of nucleons). As a consequence, the nuclear radius is very large
compared to the size of the tightly bound core. 
The large size of halo nuclei leads to threshold phenomena with
important general consequences for low-energy reaction rates in nuclear 
astrophysics. One example is the reaction $p+\mbox{$^7$Be} \to 
\mbox{$^8$B}+ \gamma$ which is important for solar neutrino production.
The nucleus $^8$B is believed to be a two-body proton halo,
consisting of a $^7$Be core and a proton \cite{BCS96}. 
Somewhat more
complicated are three-body halos consisting of a core and two slightly 
bound nucleons. Particularly interesting are Borromean three-body halos, 
where no two-body subsystem is bound. Typical examples are $^6$He 
and $^{11}$Li, which consist of a $^4$He and $^9$Li core, respectively,
and two neutrons.
For reviews of
halo nuclei,
see Ref.~\cite{ZDF93}. The physics of halo nuclei is 
an important part of the physics program at RIA \cite{RIA00}. 
A thorough
discussion of reactions with rare isotopes can be found
in Ref.~\cite{BHM02}.

The physics of halo nuclei is a promising new arena for 
Effective Field Theory (EFT).
EFTs provide a powerful framework to explore separation of scales
in physical systems in order to perform systematic, model-independent
calculations \cite{EFT}.  If, for example, the relative momentum $k$ of two
particles is much smaller than the inverse range of
their interaction $1/R$, observables can be expanded in powers of $kR$.
All short-distance effects are systematically absorbed into a few
low-energy constants using renormalization. The EFT approach allows
for systematically improvable
calculations of low-energy processes with well-defined
error estimates. The long-distance physics is included
explicitly, while the corrections from short-distance physics
are calculated in an expansion in the ratio of these two 
scales\footnote{Note that ``effective theory'' is sometimes
used in reference to a model that captures the essence
of the relevant long-distance physics without necessarily
accounting for the short-distance physics in a systematic way.
Here we use ``EFT'' in the model-independent sense described above,
in which ``power counting'' of the different orders
in the expansion is a crucial ingredient.}. 
The inherent separation of length scales in halo nuclei makes them
an ideal playing ground for EFT. 

In recent years, there has been much interest in applying
EFT methods to nuclear systems 
\cite{Birareview,NN99}.
Up to now, nuclear EFT has
mainly been applied to two-, three-, and four-nucleon systems 
starting from a fundamental nucleon-nucleon interaction.
The original motivation was to understand the gross features 
of nuclear systems from a QCD perspective by
deriving the 
nuclear potential and currents relevant for
momenta comparable to the pion mass ($p \sim m_\pi$)
\cite{Weinberg}.
More recently, it has been realized that
it is possible to carry out  
very precise calculations for fundamental physics processes
at lower energies.
For very low momenta ($p \simle m_\pi$), even pion
exchange can be considered \lq\lq short-distance'' physics. 
In this case,
one can use an effective Lagrangian including only contact interactions.
The large $s$-wave scattering lengths require that the leading two-body
contact interaction be treated nonperturbatively \cite{vKo99,KSW98}.
In the two-nucleon system, this program has been very successful
(see, {\it e.g.}, Refs.~\cite{CRS99,Birareview} and references therein).   

Using EFT, one can relate low-energy measurements in one
reaction to observables in a similar (but unmeasured) reaction
in a controlled expansion with reliable error estimates.
This is in contrast to standard potential model calculations
where errors can only be estimated by comparing different potentials.
An example of a precise calculation in the ``pionless'' EFT is
the reaction $n+p\to d+\gamma$ \cite{ChS99},
which is relevant to 
big-bang nucleosynthesis (BBN).
As for many other reactions of astrophysical interest,
the uncertainty in the cross section is difficult to determine
due to the lack of data at low energies and the lack
of information about theoretical estimates.
In the energy of relevance to BBN, both $E1$ and $M1$
capture are important.
They have been calculated to fifth and third order,
respectively, where two new counterterms appear.
Using the measured cold-capture rate and 
data for the deuteron photodisintegration reaction
to fix the counterterms,
the $n+p\to d+\gamma$ cross section was computed to 1\% for center-of-mass
energies $E\simle 1$ MeV.

Much of the strength of EFT lies in the fact that it
can be applied without
off-shell ambiguities to systems with more nucleons.
The crucial issue of the relative size of three-body forces
has been investigated in the three-body system \cite{BHK99}.
Nucleon-deuteron scattering in all channels except the $s_{1/2}$ wave
can be calculated to high orders using two-nucleon input
only, with results in striking agreement with data and
potential-model calculations \cite{BeK98}.
For example, to third order, the 
$s_{3/2}$ scattering length is found to be 
$a_{3/2}^{(EFT)}= 6.33\pm 0.10$ fm,
to be compared to the measured
$a_{3/2}^{(expt)}= 6.35\pm 0.02$ fm.
In contrast, in the $s_{1/2}$-wave channel, the non-perturbative running
of the renormalization group requires a momentum-independent
three-body force in leading order \cite{BHK00}.
Once the new parameter is fitted to (say) the scattering length,
the energy dependence is predicted.
The triton binding energy, for example, is found to be
$B_3^{(EFT)}=8.0$ MeV in leading order,
already pretty close to the experimental 
$B_3^{(expt)}=8.5$ MeV. Recently, this approach was also applied to
$\Lambda d$ scattering and the hypertriton \cite{Ham01}. 
Using the hypertriton binding energy to fix the three-body force, the 
low-energy $\Lambda d$ scattering observables can be predicted.
The results are very insensitive to the poorly known $\Lambda N$ low-energy
parameters. 
In a related study, the $\Lambda d$ Phillips line was 
established \cite{fedorovjensen}.

However, in an EFT it is by no means necessary to start from a
fundamental nucleon-nucleon interaction. If, as in halo nuclei, the 
core is much more tightly bound than the remaining nucleons,
it can be treated as an explicit degree of freedom.
One can write an EFT for the 
contact interactions of the nucleons with the
core and include the substructure of the core perturbatively
in a controlled expansion. This approach is appropriate for
energies smaller than the excitation energy of the core.
In other words, one can account for the spatial extension of the core 
by treating it as a point particle with corrections
from its finite size entering in a derivative expansion of the interaction. 
This is a consequence of the limited resolution of a long wavelength probe 
which cannot distinguish between a point and an extended particle
of size $R$ if the wavelength $\lambda \gg R$.

In this paper, we consider the virtual $p$-wave state in
$n\alpha$ scattering as a test case. Even though there is
no bound state in this channel, it has all the characteristics
of a two-body halo nucleus. 
Furthermore it is relevant for the study
of the Borromean three-body halo $^6$He,
which will be addressed in a
forthcoming publication \cite{ref6he}. 
Elastic $n\alpha$ scattering is relatively well known experimentally.
Since the nucleon has $j=1/2$ and the $\alpha$ particle has
$j=0$, there are contributions from an $s$ wave ($s_{1/2}$), 
two $p$ waves ($p_{1/2}$ and $p_{3/2}$), {\it etc.}.
Arndt, Long, and Roper performed a phase-shift analysis of low-energy
data and
extracted the effective range parameters in the $s$ and
$p$ waves \cite{ALR73}. The $p_{3/2}$ partial wave displays a resonance
at $E\sim 1$ MeV
corresponding to a shallow virtual bound state, while the 
$s_{1/2}$ and $p_{1/2}$ partial waves are nonresonant
at low energies.
We will show  that this $p$-wave resonance leads to a power counting different 
from the one for $s$-wave bound states \cite{vKo99,KSW98} that 
has been discussed extensively in the literature
because of its relevance for the no-core EFT.\footnote{
By no-core EFT we mean an EFT where all nuclei are 
dynamically generated from nucleon (and possibly pion and delta
isobar) degrees of freedom.}
In particular, proper renormalization requires two low-energy 
parameters at leading order, namely the scattering length and the 
effective range.
The extension to higher-orders is straightforward.
As we will see, the EFT describes the low-energy data very well.

The organization of this paper is as follows:
In the next Section, we work out renormalization
and power counting for a $p$-wave resonance 
in the simpler context of spinless fermions. 
In Section \ref{sec:3bdy},
we include the spin and isospin of the nucleon and
apply our formalism to elastic $n\alpha$ scattering.
In Section \ref{sec:conc}, we summarize our results
and present an outlook. In particular, we discuss the extension to 
the Borromean three-body halo $^6$He and the reaction 
$p+\mbox{$^7$Be} \to \mbox{$^8$B}+ \gamma$.

\section{EFT for Shallow $P$-Wave States}
\label{sec:pwave}

In this section, we develop the power counting for shallow $p$-wave states
(bound states or virtual states)
in the particularly simple case of
a hypothetical system of two
spinless fermions of common mass $m$. 
Our arguments are a generalization of those in Ref.~\cite{vKo99}.

In order to have a shallow bound state, we need at least
two momentum scales: the breakdown scale of the EFT, $\Mhi$,
and a second scale, $\Mlo\ll \Mhi$, that characterizes the
shallow bound state. 
The scale $\Mhi$ is set by the degrees of freedom that have been 
integrated out.
In the case of an EFT without explicit
pions and core excitations, $\Mhi$ is the smallest between 
the pion mass $m_\pi$ and the momentum corresponding
to the energy of the first excited state.
The scale $\Mlo$ is not a fundamental
scale of the underlying theory. It can be understood as arising from
a fine tuning of the parameters in the underlying theory. If the
values of these parameters were changed slightly, the scale $\Mlo$
would disappear. 
We seek an ordering of contributions at the scale $\Mlo$ 
in powers of $\Mlo/\Mhi$.
Due to the presence of fine-tuning, naive dimensional
analysis cannot be applied. 

For simplicity we neglect relativistic corrections.
They are generically small because they are suppressed by
powers of the particle mass $m$, and
in the cases of interest here $m\gg \Mhi$.
They can be included along the lines detailed in Ref.~\cite{vKo99}.

\subsection{Natural Case}
\label{natcase}

First, we will consider the natural case without any fine-tuning.
The scale of all low-energy parameters is then set by $\Mhi$
and naive dimensional analysis can be applied.

The $T$-matrix for the non-relativistic
scattering of two spinless fermions with mass $m$
in the center-of-mass frame can be expanded in partial waves as 
\beq
\label{pwT}
T(k, \cos\theta)=\sum_{l \geq 0}T_l(k, \cos\theta)=
\frac{4\pi}{m}\sum_{l \geq 0} \frac{2l+1}{k\cot\delta_l-ik}
P_l(\cos\theta)\,,
\eeq 
where $k$ is the center-of-mass momentum, $\theta$ the scattering angle,
and $P_l(\cos\theta)$ is a Legendre polynomial.\footnote{Note that we
assume the two fermions are distinguishable.}
The generalized effective range expansion for arbitrary angular
momentum $l$ reads:
\beq
\label{general_ert}
k^{2l+1}\cot\delta_l=-\frac{1}{a_l}+\frac{r_l}{2}k^2-\frac{{\cal P}_l}{4}k^4
+\ldots\,,
\eeq
where $a_l$, $r_l$, and ${\cal P}_l$ are the scattering length, 
effective range, and shape parameter in the $l$-th partial wave, 
respectively. For $l=0$, Eq.~(\ref{general_ert})
reproduces the familiar effective range expansion for
$s$ waves. 
Note that the dimension of the effective-range parameters
depends on the partial wave. 
In the $s$ wave, $a_0$ and $r_0$ have dimensions
of length, while ${\cal P}_0$ has dimensions of (length)$^3$. 
For $p$ waves,
$a_1$ has dimension of (length)$^3$
(it is a scattering ``volume''), $r_1$ has dimension 1/(length)
(it is an ``effective  momentum''), and
${\cal P}_1$ has dimension of length.

The $s$-wave contribution $T_0$ has been discussed in detail in the
literature \cite{vKo99,KSW98}.
Our goal here is to set up an EFT that reproduces the $p$-wave 
contribution $T_1$ in a low-momentum expansion,
\beq
\label{leT1}
T_1(k, \cos\theta)=-\frac{12\pi a_1}{m}k^2\cos\theta \left(1+\frac{a_1 r_1}{2}
k^2 -ia_1 k^3 +\frac{a_1}{4}(a_1 r_1^2-{\cal P}_1)k^4 +\ldots \right)\,.
 \eeq

We start with the most general Lagrangian for spinless fermions 
with $p$-wave interactions:
\beq
  {\cal L}  = \psi^\dagger \biggl[i\partial_0 + \frac{\nab^{\,2}}{2m}\biggr]
   \psi + \frac{C_2^p}{8} (\psi \galnab_i \psi)^\dagger
             (\psi\galnab_i \psi) -\frac{C_4^p}{64} \left[(\psi \galnab^{\,2}
       \,\galnab_i \psi)^\dagger (\psi\galnab_i \psi)+{\rm H.c.}
       \right]+ \ldots \, ,
  \label{lag}
\eeq      
where 
$\galnab=\overleftarrow{\nabla}-\nab$ is the Galilean invariant
derivative, H.c.\ denotes the Hermitian conjugate,
and the dots denote higher-derivative interactions 
that are suppressed at low energies.
The fermion propagator is simply
\beq
\label{ppro}
iS(p_0,\vec{p})=\frac{i}{p_0-\vec{p}^{2}/2m +i\epsilon}\,,
\eeq
and the Feynman rules for the vertices can be read off
Eq.~(\ref{lag}). 
Around the non-relativistic limit,
all interaction coefficients contain a common factor 
of $1/m$ that follows from Galilean invariance.
>From dimensional analysis, we have $C_2^p \sim 12\pi/m\Mhi^3$
and $C_4^p \sim 12\pi/m\Mhi^5$. The exact relation of 
$C_2^p$ and $C_4^p$ to the scattering length and effective range  
will be obtained in the end from matching to Eq.~(\ref{leT1}).

\begin{figure}[tb]
\begin{center}
\includegraphics[width=5in,angle=0,clip=true]{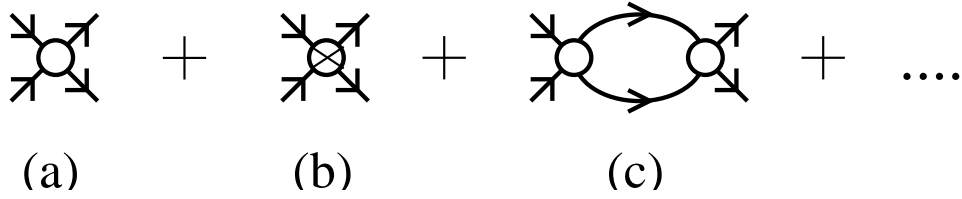}
\end{center}
\vspace*{0pt} 
\caption{Lowest-order diagrams for the perturbative expansion of $T_1$.
 Particle propagators are represented by solid lines.
 The $C_2^p\,(C_4^p)$ vertex is indicated by the open circle
 (crossed circle).}
\label{fig:pwavebub}
\end{figure}

We work in the center-of-mass frame and assign the momenta $\pm\vec{k}$
and $\pm\vec{k}'$ to the incoming and outgoing particles, 
respectively. The total energy is $E=k^2/m=k'^2/m$.
The EFT expansion is in powers of $k/\Mhi$.
The leading contribution to $T_1$ is of order $12\pi k^2/m\Mhi^3$.
It is given by the tree-level diagram with the $C_2^p$ interaction 
shown in Fig.~\ref{fig:pwavebub}(a). 
%%%%%%%%%%%%%%%%%%%%%%%%%%%%%%%%%%%%%%%%%%%%%%%%%%%%%%%%%% 
%%%%%%%%%%%%%%%%%%%%%%%%%%%%%%%%%%%%%%%%%%%%%%%%%%%%%%%%%%%%%%
The result is simply
\beq
i T_{1(a)} =-i C_2^p \,\vec{k}\cdot \vec{k}'\,. \label{T1a}
\eeq
The second term in the low-momentum 
expansion is suppressed by $k^2/\Mhi^2$ compared to the leading 
order. It is given by the tree-level diagram with the $C_4^p$ 
interaction shown in Fig.~\ref{fig:pwavebub}(b):
\beq
i T_{1(b)} =-i C_4^p k^2\,\vec{k}\cdot \vec{k}'\,. \label{T1b}
\eeq
At order $12\pi k^5/m\Mhi^6$, we have the one-loop diagram with two 
$C_2^p$ interactions shown in Fig.~\ref{fig:pwavebub}(c).
The contribution of this diagram is
\bea
iT_{1(c)} &=&(-i C_2^p)^2 \int \frac{d^4 q}{(2\pi)^4}
\frac{i\,\vec{q}\cdot \vec{k}}{\frac{E}{2}+q_0-\frac{\vec{q}^2}{2m}
+i\epsilon} \frac{i\,\vec{q}\cdot \vec{k}'}{\frac{E}{2}-q_0
-\frac{\vec{q}^2}{2m} +i\epsilon}\nonumber \\
&=& (C_2^p)^2 im k'_i k_j \int \frac{d^{3} q}{(2\pi)^{3}}
\frac{q_i q_j}{q^2-k^2-i\epsilon} \,. \label{T1c}
\eea
where the $dq_0$ integral was performed via contour integration.
The remaining integral must be proportional 
to $\delta_{ij}$ since no other vectors are available. 
Adding and subtracting $k^2$ in the numerator, we find
\bea
iT_{1(c)} &=& (C_2^p)^2 \frac{im}{6\pi^{2}} \,\vec{k}\cdot \vec{k}'\,
\left\{\int dq\, q^{2} +k^2 \int dq
+k^4\int dq\frac{1}{q^2-k^2-i\epsilon} \right\}\,,\nonumber\\
&=& (C_2^p)^2 \frac{im}{6\pi^{2}} \,\vec{k}\cdot \vec{k}'\,
\left\{ L_3 +k^2 L_1 +\frac{\pi}{2} ik^3 \right\}\,,
\label{eq:t1c}
\eea
where $L_3$ and $L_1$ are infinite constants.
These two ultraviolet divergent terms can be absorbed by redefining the 
low-energy constants $C_2^p$ and $C_4^p$, respectively, 
which are already present
in $T_{1(a)}$ and $T_{1(b)}$. No new parameter enters
at this order. 
The series proceeds in an obvious way.

We can now match to Eq.~(\ref{leT1}) to relate
the renormalized coefficients to the effective-range parameters.
We find from Eq.~(\ref{T1a}) that
$C_2^p=12\pi a_1/m$,
and from Eq.~(\ref{T1b}) that $C_4^p=C_2^p r_1 a_1/2$.
After renormalization, $T_{1(c)}$ reproduces the third 
term in the low-momentum expansion of Eq.~(\ref{leT1}). 
Note that diagram \ref{fig:pwavebub}(c) cannot be renormalized
by $C_2^p$ alone even though it does not contain a $C_4^p$
vertex. This observation has important consequences in the unnatural
case with fine-tuning.

\subsection{Unnatural Case}

Now we turn to the more interesting case with a shallow $p$-wave state.
In Refs.~\cite{vKo99,KSW98}, it was shown that for a shallow $s$-wave
state the leading-order contact interaction $C_0$ has to be treated 
nonperturbatively. In this case, $C_0$ is enhanced by a factor 
$\Mhi/\Mlo$ over the expectation $C_0\sim 4\pi/m\Mhi$ from naive 
dimensional analysis. 
Adding a new rung in the ladder forming the amplitude
means adding an intermediate state ($\sim mk/4\pi$)
and a $C_0$ ($\sim4\pi/m\Mlo$).
Since the physics of the
bound state is determined by $k\sim \Mlo$, $C_0$ 
has to be summed to all orders.

For $p$ waves matters are slightly more complicated. We have
seen above that the renormalization of the one-loop diagram with two
$C_2^p$ interactions requires tree-level counterterms
corresponding to both the leading $C_2^p$ and subleading $C_4^p$
interaction. As consequence, at least the $C_2^p$ and $C_4^p$ interactions 
have to be treated nonperturbatively if a shallow $p$-wave state is
present. 

Bound (virtual) states are associated with 
poles in the $S$-matrix
on the upper (lower) half of the 
complex momentum plane.
The characteristic momentum $\gamma$ of the bound/virtual state
is given by position of the pole, $|k|\equiv \gamma$. 
For a shallow 
$p$-wave state with $\gamma\sim\Mlo$, the magnitude of both the effective 
range and the scattering length must be set by $\Mlo$.
This is a consequence of the renormalization argument from the previous
subsection. Either $C_2^p$ and $C_4^p$ are both
enhanced or they are both natural. Assuming that the higher terms
in the effective range expansion are natural,
the order of magnitude of the first three terms in the expansion is
\beq
k^3\cot\delta_1 \sim\Mlo^3+\Mlo k^2 +\frac{1}{\Mhi}k^4+\ldots\,,
\eeq
and the effective-range parameters scale as
\beq \label{ordermag}
\frac{1}{a_1}\sim\Mlo^3\,,\quad \frac{r_1}{2}\sim\Mlo\,,\quad\mbox{ and }\quad
\frac{{\cal P}_1}{4}\sim\frac{1}{\Mhi}\,.
\eeq
Both $C_2^p$ and $C_4^p$ are enhanced over 
the expectation from naive dimensional analysis and scale as
\beq
C_2^p \sim \frac{12\pi}{m\Mlo^3}\qquad \mbox{and}
\qquad C_4^p\sim \frac{12\pi}{m\Mlo^5}\,.
\eeq
Consequently, for momenta of order $\Mlo$ neither interaction can be 
treated perturbatively. The shape parameter ${\cal P}_1$, however, is of 
order $1/\Mhi$ and its contribution is suppressed by $\Mlo/\Mhi$
compared to the leading order. 

In the following, we will demonstrate that treating
the $C_2^p$ and $C_4^p$ interactions to all orders is indeed sufficient 
for proper renormalization and, moreover, required to reproduce the 
physics of the shallow $p$-wave state.
We will also work out the leading-order description
of a shallow $p$-wave state. 

For convenience, we will not use the Lagrangian (\ref{lag}) but
follow Ref.~\cite{Kap97} and introduce an auxiliary field 
(the dimeron) for the  two-particle state. The corresponding Lagrangian is,
\bea
\label{lagd}
{\cal L}&=&\psi^\dagger \biggl[i\partial_0 + \frac{\nab^{\,2}}{2m}\bigg]
   \psi+ \eta_1 d^\dagger_i\bigg( i\partial_0+\frac{\nab^{\,2}}{4m}
-\Delta_1 \bigg)d_i +\frac{g_1}{4}\left(d^\dagger_i (\psi \galnab_i \psi) +
{\rm H.c.}\right)+\ldots\,,
\eea
where the sign $\eta_1=\pm 1$ and the parameters $g_1$ and $\Delta_1$ will
be fixed from matching.
This Lagrangian contains exactly the same number of parameters as
the original Lagrangian (\ref{lag}). Up to higher order terms,
Eq.~(\ref{lagd}) is equivalent to Eq.~(\ref{lag}), as can be 
seen by performing the Gaussian path integral over $d_i$.

\begin{figure}[tb]
\begin{center}
\includegraphics[width=5in,angle=0,clip=true]{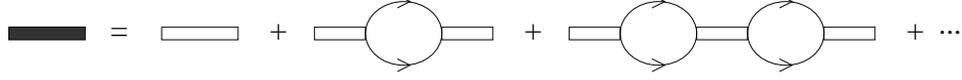}
\end{center}
\vspace*{-0pt} 
\caption{The full dimeron propagator (thick shaded line) is
obtained by dressing the bare dimeron propagator (double solid
line) with particle bubbles (solid lines) to all orders.}
\label{fig:auxprop}
\end{figure}

The bare dimeron propagator is given by
\beq
\label{dbare}
iD_{1}^0(p_0,\vec{p})_{ij}=\frac{i\eta_1\delta_{ij}}{p_0-\vec{p}^{2}/4m
-\Delta_1 +i\epsilon}\,.
\eeq
Summing the $C_2^p$ and $C_4^p$ interactions to all orders
in the theory without the dimeron corresponds to dressing
the bare dimeron propagator with particle bubbles to all orders.
This summation is shown diagrammatically in Fig.~\ref{fig:auxprop}.
%%%%%%%%%%%%%%%%%%%%%%%%%%%%%%%%%%%%%%%%%%%%%%%%%%%%%%%%%% 
%%%%%%%%%%%%%%%%%%%%%%%%%%%%%%%%%%%%%%%%%%%%%%%%%%%%%%%%%%%%%%
The full dimeron propagator is most easily calculated by first
computing the self-energy from the particle bubble $-i\Sigma_{1}$
which up to overall factors is given by the one-loop diagram from
Fig.~\ref{fig:pwavebub}(c). We have
\bea
-i\Sigma_1 (p_0,\vec{p})_{ij}&=& g_1^2 \int \frac{d^4 q}{(2\pi)^4}
\frac{q_i q_j}{\left(\frac{p_0}{2}+q_0-\frac{(\vec{p}/2+\vec{q})^2}{2m}
+i\epsilon\right)\left(\frac{p_0}{2}-q_0-\frac{(\vec{p}/2-\vec{q})^2}{2m} 
+i\epsilon\right)}\nonumber \\
&=& i\delta_{ij} \frac{m g_1^2}{12\pi} \left\{
\frac{2}{\pi}L_3+\frac{2}{\pi}L_1 \left(mp_0-\vec{p}^{2}/4 \right)
+i \left(mp_0-\vec{p}^{2}/4 \right)^{3/2} \right\},
\label{sigma}
\eea
where $L_3$ and $L_1$ are infinite constants as in Eq.~(\ref{eq:t1c}).
The full dimeron propagator now simply follows from the geometric series
\bea
iD_1(p_0,\vec{p})&=&iD_1^0(p_0,\vec{p})+iD_1^0(p_0,\vec{p})
(-i\Sigma_1(p_0,\vec{p})) iD_1^0(p_0,\vec{p})+\ldots\nonumber\\
&=&iD_1^0(p_0,\vec{p})\left(1-\Sigma_1(p_0,\vec{p})
D_1^0(p_0,\vec{p})\right)^{-1}\,,
\label{geo}
\eea
where the vector indices have been suppressed. Using Eqs.~(\ref{dbare}, 
\ref{sigma}), we find
\bea
iD_1 (p_0,\vec{p})_{ij}&=& -i\delta_{ij}\frac{12\pi}{mg_1^2}
\left( \frac{12\pi\Delta_1}{\eta_1 mg_1^2} 
-\frac{12\pi}{\eta_1 m^2 g_1^2}\left(mp_0-\vec{p}^2/4 \right)\right. \\
& & \qquad\qquad\left. -\frac{2}{\pi}L_3 
                 -\frac{2}{\pi}L_1 \left(mp_0-\vec{p}^2/4\right)
-i \left(mp_0-\vec{p}^{2}/4\right)^{3/2} \right)^{-1}\,,\nonumber\\
&=& -i\delta_{ij}\frac{12\pi}{mg_1^2}
\left( \eta_1 \frac{12\pi\Delta_1^R}{m(g_1^R)^2} 
-\eta_1 \frac{12\pi}{m^2 (g_1^R)^2}\left(mp_0-\vec{p}^2/4 \right)
-i \left(mp_0-\vec{p}^{2}/4\right)^{3/2} \right)^{-1}\,,\nonumber
\eea
where the last line defines the renormalized parameters
$\Delta_1^R$ and $g_1^R$.

The $p$-wave scattering amplitude is
obtained by attaching 
external particles lines to the full dimeron propagator.
In the center-of-mass system, $(p_0, \vec{p})=(k^2/m,\vec{0})$, 
this leads to
\bea \label{T_1}
T_1(k,\cos\theta)&=&\frac{12\pi}{m}k^2\cos\theta
\left( \eta_1 \frac{12\pi\Delta_1}{m (g_1^R)^2}
      -\eta_1 \frac{12\pi}{m^2 (g_1^R)^2}k^2-ik^3 \right)^{-1}
\nonumber\\
&\equiv &\frac{12\pi}{m}k^2\cos\theta\left(-\frac{1}{a_1}+\frac{r_1}
{2}k^2 -ik^3 \right)^{-1}\,,
\eea
from which the matching conditions can be read off easily.
We see that, as advertised, two coefficients
are necessary and sufficient to remove 
any significant cutoff dependence.

\subsection{Pole Structure}
\label{polestruc}

In this subsection, we discuss the pole structure of the $S$-matrix
in the unnatural case.
Neglecting terms suppressed by $\Mlo/\Mhi$,
the equation determining the poles is,
from the amplitude (\ref{T_1}),
\beq
-\frac{1}{a_1}+\frac{r_1}{2} \kappa^2 -i\kappa^3 =0\,.
\eeq

\begin{figure}[tb]
\begin{center}
\includegraphics[width=3in,angle=0,clip=true]{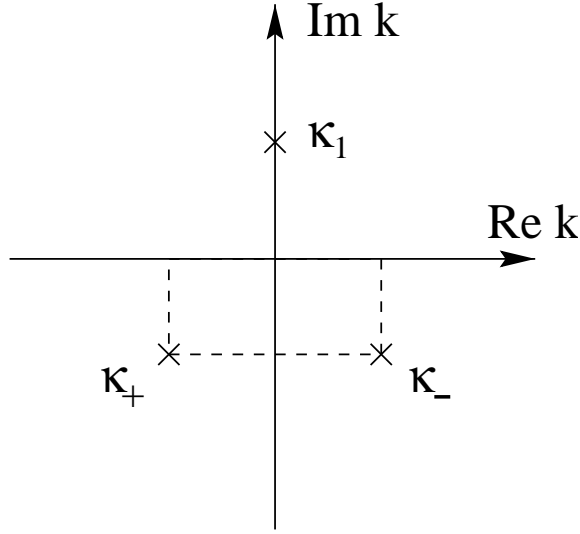}
\end{center}
\vspace*{-0pt} 
\caption{The pole structure of the $S$-matrix for a $p$-wave resonance.}
\label{fig:ppoles}
\end{figure}

For definiteness,
we concentrate on the case $a_1, r_1 <0$ that is relevant to 
$n\alpha$ scattering. Other cases can be examined as easily.
The solutions are one pole $\kappa_1$ on the positive imaginary axis
and two complex-conjugated poles in the lower half-plane. They have
the structure
\beq
\kappa_1 = i \gamma_1 \qquad\mbox{and}\qquad
\kappa_\pm = i( \gamma \pm i\tilde{\gamma})\,,
\eeq
where 
\bea
\gamma_1 &=&\frac{1}{6}\left( |r_1| +\frac{|a_1|^{1/3} |r_1|^2}{v}
         +\frac{v}{|a_1|^{1/3}} \right) \,,\nonumber\\
\gamma &=&\frac{1}{6}\left( |r_1| -\frac{|a_1|^{1/3} |r_1|^2}{2v}
         -\frac{v}{2|a_1|^{1/3}} \right) \,,\nonumber\\
\tilde{\gamma} &=&-\frac{\sqrt{3}}{12}\left(\frac{|a_1|^{1/3} |r_1|^2}{v}
         -\frac{v}{|a_1|^{1/3}} \right) \,,\nonumber\\
v&=&\left(108+|a_1||r_1|^3 +108\sqrt{1+|a_1||r_1|^3/54}\right)^{1/3}
\,.
\eea
This pole structure is illustrated in Fig.~\ref{fig:ppoles}.
%%%%%%%%%%%%%%%%%%%%%%%%%%%%%%%%%%%%%%%%%%%%%%%%%%%%%%%%%% 
%%%%%%%%%%%%%%%%%%%%%%%%%%%%%%%%%%%%%%%%%%%%%%%%%%%%%%%%%%%%%%
This general structure remains qualitatively unchanged in the limit 
$|r_1|\to 0$. 

The $p$-wave contribution to the $S$-matrix can be 
written as
\beq
S_1=e^{2i\delta_1}=-\frac{k+\kappa_1}{k-\kappa_1}\,
 \frac{k+\kappa_+}{k-\kappa_+}\,\frac{k+\kappa_-}{k-\kappa_-}
=-\frac{k+i\gamma_1}{k-i\gamma_1}\,\frac{E-E_0-\frac{i}{2}\Gamma(E)}
{E-E_0+\frac{i}{2}\Gamma(E)}\,,
\eeq
where we have defined
\beq
\label{res_defs}
E=\frac{k^2}{2\mu}\,,\qquad
E_0=\frac{\gamma^2+\tilde{\gamma}^2}{2\mu}
\qquad\mbox{and}\qquad \Gamma(E)=-4\gamma \sqrt{\frac{E}{2\mu}}\,,
\eeq
with $\mu$ the reduced mass of the system.
The phase shift can therefore be written as
\beq
\delta_1=\frac{1}{2i}\ln S_1=
         \delta_s(E)-\arctan\left(\frac{\Gamma(E)}{2(E-E_0)}\right)\,.
\label{eq:pphase}
\eeq
Here 
\beq
\delta_s(E) =\frac{1}{2} \arctan \left(\frac{2\sqrt{EB}}{E-B}\right)
\eeq
is the contribution from the bound state with binding energy
$B=\gamma_1^2/2\mu$.
It changes by $\pi/2$ as the energy varies across $B$.
$\delta_s(E)$ is a relatively smooth function of the energy $E$.
The two complex-conjugated poles $\kappa_\pm$ generate the 
resonance that is given by the second term in Eq.~(\ref{eq:pphase}).
This term changes by $\pi$ as the energy varies across $E_0$.

In the case of $s$ waves, the EFT 
determines in leading order the position of a shallow real or virtual bound
state. 
In the $p$ waves the physics is richer: 
the two leading-order parameters provide 
the position and width of a resonance
(in addition to the position of a bound state).

\section{Application to Elastic $n\alpha$ Scattering}
\label{sec:3bdy}

We are now in position to extend the EFT for shallow $p$-wave states from the
previous section 
to the $n$-$^4$He system, including the spin of the nucleon. 
We 
calculate the leading- and next-to-leading-order contributions
to low-energy elastic $n\alpha$ scattering. 
First, we briefly review
the structure of the cross section and scattering amplitude.

\subsection{Cross Section and Scattering Amplitude}

The differential cross section for elastic $n\alpha$ scattering
in the center-of-mass frame can be written as
\beq
\frac{d\sigma}{d\Omega}=|F(k,\theta)|^2+|G(k,\theta)|^2\,,
\eeq
where $k$ and $\theta$ are the magnitude of the momentum and 
the scattering angle, respectively.
The so-called spin-no-flip and spin-flip amplitudes $F$ and $G$ can
be expanded in partial waves as
\bea
F(k,\theta)&=&\sum_{l \geq 0} \left[ (l+1) f_{l+}(k) +l f_{l-}(k)
      \right]P_l (\cos\theta)
\,,\label{eq:defF}\\
G(k,\theta)&=&\sum_{l \geq 1} \left[ f_{l+}(k)-f_{l-}(k) \right]
      P^1_l (\cos\theta)\,,
\label{eq:defG}
\eea
where $P_l$ is a Legendre polynomial and
\beq
\label{eq:defP1}
P_l^1(x)=(1-x^2)^{1/2} \frac{d}{dx} P_l(x)\,.
\eeq
The partial wave amplitudes $f_{l\pm}$ are related to the 
phase shifts $\delta_{l\pm}$ via
\beq
\label{eq:napartial}
f_{l\pm}(k)=\frac{1}{2ik}\left[ e^{2i\delta_{\pm}}-1 \right]
=\frac{1}{k\cot\delta_{l\pm}-ik}\,.
\eeq
The total cross section can be obtained from the optical theorem,
\beq
\label{eq:sigtot}
\sigma_T=\frac{4\pi}{k}\,{\rm Im}\,F(k,0)\,.
\eeq
The $T$-matrix calculated in EFT is related to the amplitudes 
$F$ and $G$ via
\beq
\label{eq:defT}
T=\frac{2\pi}{\mu}\left( F+i \vecsigma\cdot\hat{\vec n} G\right)\,,
\eeq
where $\mu=m_\alpha m_N/(m_\alpha+m_N)$ is the reduced mass, 
$\hat{\vec n}=\vec{k}\times\vec{k}'/|\vec{k}\times\vec{k}'|$ with 
$\vec{k}$ and $\vec{k}'$ the initial and final momenta in the 
center-of-mass frame, and $\vecsigma=(\sigma_1,\sigma_2,\sigma_3)$ 
is a three-vector of the usual Pauli matrices.

For $n\alpha$ scattering at low energies only the $s$ and $p$ waves
are important. There is one $s$ wave: $l\pm=0+$ with $l_j=s_{1/2}$, and
two $p$ waves: $l\pm=1+$ and $1-$ corresponding to $l_j=p_{3/2}$ and 
$p_{1/2}$, respectively. In the remainder of the paper, we use
the $l\pm$ notation for the partial waves.
In Ref.~\cite{ALR73}, a phase-shift analysis including the $0+$, 
$1-$, and $1+$ partial waves was performed and the effective-range
parameters were extracted.
The effective range expansion for a partial wave
with orbital angular momentum $l$ was given in Eq.~(\ref{general_ert}).
The effective-range parameters extracted in Ref.~\cite{ALR73} are
listed in Table~\ref{tab1}.     
%%%%%%%%%%%%%%%%%%%%%%%%%%%%%%%%%%%%%%%%%%%%%%%%%%%%%%%%%%%%%%%%%%%%%%%%
\begin{table}[b]
\begin{tabular}{c||c|c|c}
Partial wave $l_{\pm}$ & $a_{l\pm}$ [fm$^{1+2l}$] & $r_{l\pm}$ 
[fm$^{1-2l}$] &  ${\cal P}_{l\pm}$ [fm$^{3-2l}$]
\\ \hline\hline
$0+$ & $2.4641(37)$ & $1.385(41)$ & $-$ \\
$1-$ & $-13.821(68)$ & $-0.419(16)$ & $-$\\
$1+$  & $-62.951(3)$ & $-0.8819(11)$ & $-3.002(62)$
\end{tabular}
\caption{The values of the scattering length $a_{l\pm}$, the effective 
range $r_{l\pm}$, and the shape parameter ${\cal P}_{l\pm}$ 
in elastic $n\alpha$ scattering for the $0+$, $1-$, and $1+$ partial 
waves from Ref.~\protect\cite{ALR73}. The numbers in parenthesis indicate
the error in the last quoted digits. All values are given in units of the
appropriate powers of fm as determined by the orbital angular momentum $l$
of the partial wave.}
\label{tab1}
\end{table}
%%%%%%%%%%%%%%%%%%%%%%%%%%%%%%%%%%%%%%%%%%%%%%%%%%%%%%%%%%%%%%%%%%%%%%%% 
The $1+$ partial wave has a large scattering length and
somewhat small effective range, as expected from Eq.~(\ref{ordermag}).
Indeed,
the phase shift
in this wave has a resonance corresponding to a shallow
$p$-wave state \cite{ALR73}. 
As a consequence, the $1+$ partial wave has to be treated 
nonperturbatively using the formalism for shallow $p$-wave
states developed in the previous section. 
In the $0+$ wave, on the other hand, the scattering
length and effective range are clearly of natural size.
The $0+$ partial wave can be treated in perturbation theory.
The situation is less clear in the $1-$ wave.
Although the pattern is similar to the $1+$ wave,
the phase shifts in the $0+$ and $1-$ partial waves show no 
resonant behavior at low energies \cite{ALR73}. 
We therefore expect that perturbation theory can be applied
to the $1-$ partial wave as well. 

These points can be made slightly more precise.
We can estimate the scales $\Mlo$ and $\Mhi$ from the 
effective-range parameters.
Using the parameters for the $1+$ partial wave
from Table~\ref{tab1}, we find for $\Mlo$ 50 MeV from the scattering 
length and 90 MeV from the effective range. The average value is
$\Mlo\approx 70$ MeV. From the shape parameter, we extract 
$\Mhi\approx 260$ MeV. This is consistent with the hierarchy
$\Mlo \ll \Mhi \sim m_\pi \sim \sqrt{m_N E_\alpha}$, 
where $E_\alpha= 20.21$ MeV is 
the excitation energy of the $\alpha$ core \cite{tunl},
and suggests that our power counting
is appropriate for the $1+$ partial wave.
We would expect that the scale of all effective-range parameters
in the remaining channels is set by $\Mhi$. Extracting the numbers,
however, we find for $\Mhi$ the scales 
80 MeV from $a_{0+}$, 280 MeV from $r_{0+}$,
80 MeV from $a_{1-}$, and 40 MeV from $r_{1-}$. 
While some spread is not surprising given the qualitative nature
of the argument,
these numbers suggest that, even though the $1-$ phase shift is small,
this partial wave might also be dominated by $\Mlo$. 
For the moment we will assume this is not the case
and treat the $1-$ wave in perturbation theory.
We can certainly improve convergence by resumming $1-$ contributions.
We return to this point in Sect.~\ref{further}.

\subsection{Scattering Amplitude in the EFT}

A real test of the power counting comes only by 
calculating the amplitude at various orders
and comparing the results among themselves and with data.
In the following, we will compute $n\alpha$ scattering 
to next-to-leading order in the EFT. 
For characteristic momenta $k\sim \Mlo$,
the leading-order contribution to the $T$-matrix is
of order $12\pi/m\Mlo$. The EFT expansion is in $\Mlo/\Mhi$ 
and  the NLO and N$^2$LO contributions are suppressed
by powers of $\Mlo/\Mhi$ and $\Mlo^2/\Mhi^2$, respectively.
The parameters in the effective
Lagrangian will be determined from matching to effective-range
parameters. 
We then compare our results with the 
phase-shift analysis \cite{ALR73} 
and also directly with low-energy data. 

We represent the nucleon and the $^4$He core 
by a spinor/isospinor $N$ field 
and a scalar/isoscalar $\phi$ field, respectively.
We also introduce isospinor dimeron fields that
can be thought of as bare fields for the various
$N\alpha$ channels.
In the following we will employ 
$s$, $d$, and $t$, which
are spinor, spinor and four-spinor fields associated
with the $s_{1/2}$, $p_{1/2}$, and $p_{3/2}$ channels, 
respectively.

The parity- and time-reversal-invariant Lagrangians for LO and NLO 
are\footnote{We make a particular choice of fields here.
The $S$-matrix is independent of this choice.
One can, for example, redefine the $t$ field so as to remove the
$g_{1+}'$ term. In this case, its contribution 
(see Eq.~(\ref{eq:Tnlo}) below) is reproduced
by a $t^\dagger N\phi$ (+ H.c.) interaction with three derivatives.}
\bea
{\cal L}_{\rm LO}&=&\phi^\dagger \biggl[i\partial_0 + 
\frac{\nab^{\,2}}{2m_{\alpha}}\bigg]\phi +N^\dagger \biggl[i\partial_0
+ \frac{\nab^{\,2}}{2m_{N}}\bigg]N
+  \eta_{1+} t^\dagger\bigg[i\partial_0+\frac{\nab^{\,2}}{2(m_\alpha+m_N)}
-\Delta_{1+} \bigg] t \nonumber\\
&& +\frac{g_{1+}}{2}\bigg\{ t^\dagger \vec{S}^\dagger\cdot
\bigg[N \nab \phi-(\nab N)\phi\bigg]+ {\rm H.c.} -r\bigg[t^\dagger 
\vec{S}^\dagger\cdot\nab(N \phi)+ {\rm H.c.}\bigg]\bigg\}\,,
\label{lagLO}\\
{\cal L}_{\rm NLO}&=&\eta_{0+} s^\dagger \bigg[
         -\Delta_{0+} \bigg] s +g_{0+} \bigg[s^\dagger N \phi 
+\phi^\dagger N^\dagger s \bigg]
+  g_{1+}'  t^\dagger\bigg[i\partial_0+\frac{\nab^{\,2}}{2(m_\alpha+m_N)}
\bigg]^2 t\,,
\label{lagNLO}
\eea
where $r=(m_\alpha-m_N)/(m_\alpha+m_N)$.
The notation is analogous to that in Eq.~(\ref{lagd}). 
The $S_i$'s are the $2\times 4$ spin-transition
matrices connecting states with total angular momentum
$j=1/2$ and $j=3/2$. They satisfy the relations
\bea
&&S_i S_j^\dagger=\frac{2}{3}\delta_{ij}-\frac{i}{3}\epsilon_{ijk}\sigma_k
\nonumber\\[2pt]
&&S_i^\dagger S_j=\frac{3}{4}\delta_{ij}-\frac{1}{6}\bigg\{
J_i^{3/2},J_j^{3/2}\bigg\}+\frac{i}{3}\epsilon_{ijk}J_k^{3/2}\,,
\eea
where the $J_i^{3/2}$ are the generators of the $J=3/2$ representation of 
the rotation group, with
\beq
\bigg[J_i^{3/2},J_j^{3/2}\bigg]=i\epsilon_{ijk} J_k^{3/2}\,.
\eeq
These Lagrangians generate 
contributions in the $1+$ and $0+$ partial waves. 
There are no contributions in N$^2$LO,
and the $1-$ partial wave enters first at N$^3$LO.

The propagator for the $\phi$ field is
\beq
\label{prop4he}
iS_{\phi}(p_0,\vec{p})=\frac{i}{p_0-\vec{p}^{\,2}/2m_\alpha +i\epsilon}\,,
\eeq
while the nucleon propagator is
\beq
\label{propN}
iS_{N}(p_0,\vec{p})_{\alpha\beta}^{ab}=
\frac{i\delta_{\alpha\beta}\delta_{ab}}{p_0-\vec{p}^{\,2}/2m_N +i\epsilon}\,.
\eeq
In Eq.~(\ref{propN}), $\alpha$ and $\beta$ ($a$ and $b$) are the 
incoming and outgoing spin (isospin) indices of the nucleon, respectively.
The bare propagator for the $1+$ dimeron is
\beq
\label{d32bare}
iD^{0}_{1+} (p_0,\vec{p})_{\alpha\beta}^{ab}
=\frac{i\eta_{1+}\delta_{\alpha\beta}\delta_{ab}}{p_0-\vec{p}^{\,2}/
2(m_\alpha+m_N) -\Delta_{1+} +i\epsilon}\,,
\eeq
with $\alpha$ and $\beta$ ($a$ and $b$) the 
incoming and outgoing spin (isospin) indices of the dimeron, respectively.
Note that $\delta_{\alpha\beta}$ is a $4\times 4$ unit matrix,
since the dimeron carries $j=3/2$.
The bare propagator for the $0+$ is slightly different
because its kinetic terms do not appear until higher order:
\beq
\label{s12bare}
iD^{0}_{0+} (0,\vec{0})_{\alpha\beta}^{ab}
=-\frac{i\eta_{0+}\delta_{\alpha\beta}\delta_{ab}}{\Delta_{0+}}\,,
\eeq
with $\delta_{\alpha\beta}$ now a $2\times 2$ 
unit matrix.
The bare propagator for the $1-$ dimeron is the same as for the 
$0+$ dimeron, with the index $0+$ replaced by $1-$ where appropriate.

The leading contribution to the $n\alpha$ scattering amplitude
for $k\sim \Mlo$
is of order $12\pi/m\Mlo$ and comes solely from the $1+$ partial wave
with the scattering-length and effective-range terms included to 
all orders.
The next-to-leading order correction is suppressed by $\Mlo/\Mhi$
and fully perturbative. It consists of the correction from the shape 
parameter ${\cal P}_{1+}$ in the $1+$ partial wave and the tree-level
contribution of the scattering length $a_{0+}$ in the $0+$
partial wave. The $1-$ partial wave still vanishes at 
next-to-leading  order. 

First, we calculate the leading-order $T$-matrix element $T^{\rm LO}$.
As demonstrated for spinless fermions in the previous section,
this is most easily achieved by first calculating the full dimeron 
propagator for the $1+$ dimeron and attaching the external
particle lines in the end. Apart from the spin/isospin algebra,
the calculation is equivalent to the one for spinless fermions
that was discussed in detail in the previous section.
The proper self energy is given by
\bea
-i\Sigma_{1+} (p_0,\vec{p})_{\alpha\beta}^{ab}&=&
g_{1+}^2 \int \frac{d^d l}{(2\pi)^d}
\frac{(\vec{l}-r\vec{p}/2)_i (S_i^\dagger)_{\beta\gamma}
(\vec{l}-r \vec{p}/2)_j (S_j)_{\gamma\alpha} \delta_{ab}}
{\left(\frac{p_0}{2}+l_0-\frac{(\vec{p}/2+\vec{l})^2}{2m_\alpha}
+i\epsilon\right)\left(\frac{p_0}{2}-l_0-\frac{(\vec{p}/2-\vec{l})^2}{2m_N} 
+i\epsilon\right)}\nonumber \\
&=& -ig_{1+}^2 (S_i^\dagger S_j)_{\beta\alpha} \delta_{ba}
\int \frac{d^{d-1} l}{(2\pi)^{d-1}} \frac{(\vec{l}-r\vec{p}/2)_i
(\vec{l}-r\vec{p}/2)_j}{p_0-(p^2/4+l^2-r\vec{p}\cdot\vec{l})/2\mu
 +i\epsilon}\,,
\eea
where we have performed the $dl_0$ integral via contour integration.
Evaluating the remaining integral using dimensional regularization
with minimal subtraction for simplicity, 
we obtain
\beq
\Sigma_{1+} (p_0,\vec{p})_{\alpha\beta}^{ab}=-
\delta_{\alpha\beta}\delta_{ab}\frac{g_{1+}^2 \mu}{6\pi} 
\left[2\mu\left(-p_0+\frac{\vec{p}^{\,2}}{2(m_\alpha+m_N)}-i\epsilon
\right)\right]^{3/2}\,.
\eeq
Using Eq.~(\ref{geo}), the full dimeron propagator is then given by
\bea
iD_{1+}(p_0,\vec{p})_{\alpha\beta}^{ab}&=&{i\eta_{1+}\delta_{\alpha\beta}
\delta_{ab}}\left(p_0-\frac{\vec{p}^{\,2}}{2(m_\alpha+m_N)}-\Delta_{1+} 
\right.\nonumber\\
&&\qquad\left. +\frac{\eta_{1+} \mu g_{1+}^2}{6\pi}
(2\mu)^{3/2}\left[-p_0+\frac{\vec{p}^{\,2}}{2(m_\alpha+m_N)}-i\epsilon
\right]^{3/2} +i\epsilon\right)^{-1}\,.
\eea
The leading-order $T$-matrix element in the center-of-mass system is
obtained by setting $(p_0, \vec{p})=(k^2/2\mu,\vec{0})$ and attaching 
external particles lines to the full dimeron propagator.
This leads to
\bea
T^{\rm LO}(k,\cos\theta)&=&\frac{2\pi}{\mu}
k^2( 2 \cos\theta +i\vecsigma\cdot \hat{\vec n} \sin\theta)
\left( \eta_{1+}\frac{6\pi\Delta_{1+}}{\mu g_{1+}^2}
       -\eta_{1+}\frac{6\pi}{\mu^2 g_{1+}^2}
 \frac{k^2}{2}-ik^3 \right)^{-1}
\,.
\eea
Using Eqs. (\ref{general_ert}) and (\ref{eq:defF}) to (\ref{eq:defT}), 
we find the matching conditions
\beq\label{eq:matchlo}
a_{1+}=-\eta_{1+} \frac{\mu g_{1+}^2}{6\pi\Delta_{1+}}
\qquad\mbox{ and } \qquad
{r_{1+}}=-\eta_{1+} \frac{6\pi}{\mu^2 g_{1+}^2}\,,
\eeq
which determine the parameters $g_{1+}$, $\Delta_{1+}$, and the
sign $\eta_{1+}$ in terms of the effective-range parameters
$a_{1+}$ and $r_{1+}$.
Then,
\bea
F^{\rm LO}(k,\theta)&=&\frac{2k^2\cos\theta}{-1/a_{1+}+r_{1+}k^2/2-ik^3}\,,
\nonumber\\
G^{\rm LO}(k,\theta)&=&\frac{k^2\sin\theta}{-1/a_{1+}+r_{1+}k^2/2-ik^3}\,.
\eea

At next-to-leading order, 
we include all contributions that are suppressed by
$\Mlo/\Mhi$ compared to the leading order. These contributions
come from the shape parameter ${\cal P}_{1+}$ and the $s$-wave scattering
length $a_{0+}$. Using the Lagrangian~(\ref{lagLO}) and (\ref{lagNLO}),
we find for the $T$-matrix element:
\bea\label{eq:Tnlo}
T^{\rm NLO}&=&\frac{\eta_{0+}g_{0+}^2}{\Delta_{0+}}
+ \frac{3\pi g_{1+}'}{2\mu^3 g_{1+}^2}
  \frac{2\pi}{\mu}\frac{k^6(2 \cos\theta 
                        +i\vecsigma\cdot \hat{\vec n} \sin\theta)}
{(-1/a_{1+}+r_{1+}k^2/2-ik^3)^2} \,.
\eea
The first term in Eq.~(\ref{eq:Tnlo}) corresponds to the scattering length
in the $0+$ wave, while the second term corresponds to 
the $1+$ amplitude with the shape parameter treated as a perturbation.
Using Eqs. (\ref{general_ert}) and (\ref{eq:defF}) to (\ref{eq:defT}), 
we find
\beq
\label{eq:matchnlo}
a_{0+}=-\frac{\eta_{0+}g_{0+}^2 \mu}{2\pi\Delta_{0+}}
\qquad \mbox{ and }\quad {\cal P}_{1+}=\frac{6 \pi g_{1+}'}{\mu^3 g_{1+}^2}
\,.
\eeq
Note that to this order $\eta_{0+}$, $g_{0+}$, and $\Delta_{0+}$
are not independent and only the combination appearing in
Eq.~(\ref{eq:matchnlo}) is determined. 
The next-to-leading-order pieces of $F$ and $G$ are then
\bea
F^{\rm NLO}(k,\theta)&=&-a_{0+}+\frac{{\cal P}_{1+}}{4}
\frac{2 k^6\cos\theta}{(-1/a_{1+}
+r_{1+}k^2/2-ik^3)^2}\,,\nonumber\\
G^{\rm NLO}(k,\theta)&=&\frac{{\cal P}_{1+}}{4}
\frac{k^6\sin\theta}{(-1/a_{1+}
+r_{1+}k^2/2-ik^3)^2}\,.
\eea

\subsection{Phase Shifts and Cross Sections in the EFT}

In order to see how good our expansion is, we need to
fix our parameters.
In principle we could determine the parameters by matching
our EFT to the underlying EFT whose degrees of
freedom are nucleons (and possibly pions and delta isobars),
but no core.
Unfortunately, calculations with the latter EFT have not yet reached systems
of five nucleons \cite{Birareview}.
For the time being, we need to determine the parameters from data.
For simplicity, we use 
the effective-range parameters 
from Table~\ref{tab1} together with
Eqs. (\ref{eq:matchlo},\ref{eq:matchnlo}).

\begin{figure}[tb]
\begin{center}
\includegraphics[width=4in,angle=0,clip=true]{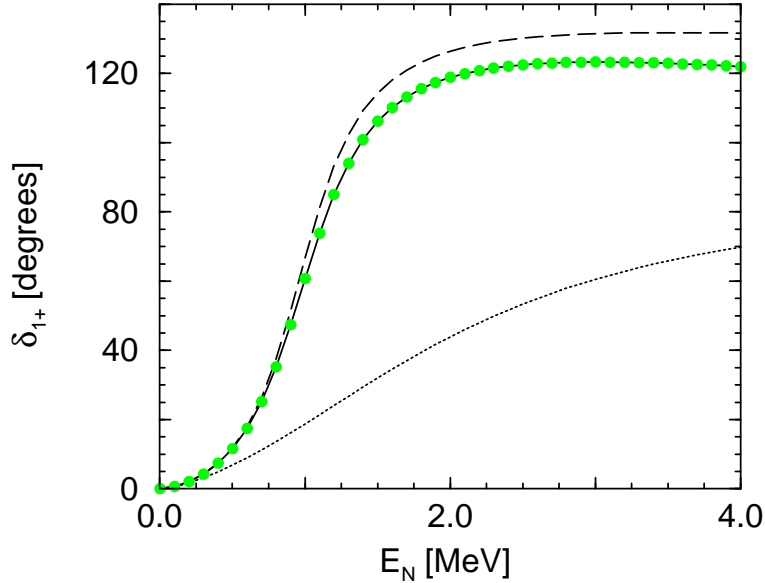}
\end{center}
\vspace*{-0pt} 
\caption{The phase shift for $n\alpha$ scattering in the 
 $1+$ partial wave as a function of the neutron kinetic energy
 in the $\alpha$ rest frame. The dashed (solid) line shows the EFT result 
 at LO (NLO). The filled circles show the phase-shift analysis 
 \protect\cite{ALR73} which the EFT at NLO reproduces exactly. 
 The dotted line shows the contribution of the scattering length
 alone.}
\label{fig:nhe4P32}
\end{figure}

In Fig.~\ref{fig:nhe4P32}, we show the phase shifts for
elastic $n\alpha$ scattering in the $1+$ partial wave
as a function of the neutron kinetic energy
in the $\alpha$ rest frame.
%%%%%%%%%%%%%%%%%%%%%%%%%%%%%%%%%%%%%%%%%%%%%%%%%%%%%%%%%% 
%%%%%%%%%%%%%%%%%%%%%%%%%%%%%%%%%%%%%%%%%%%%%%%%%%%%%%%%%%%%%%
The filled circles show the phase-shift analysis of
Ref.~\cite{ALR73}. The dashed line shows the EFT result at leading
order. 
The LO result already shows a good agreement with the
full phase-shift analysis.
As expected, the agreement deteriorates with energy.
NLO corrections improve the agreement:
the EFT result at NLO shown by the solid line
reproduces the phase-shift analysis exactly.
If better data were available and a more complete
phase-shift analysis were performed, some small discrepancies
would survive, to be remedied by higher orders.

The sharp rise in the $1+$ phase shift past $\pi/2$ denotes
the presence of a resonance.
To LO, the pole structure of the $S$-matrix is given
in Sect. \ref{polestruc}. We find
$\gamma_1 = 99$ MeV, 
$\gamma = -6$ MeV, and 
$\tilde{\gamma} = 34$ MeV.
Using Eq.~(\ref{res_defs}), the position and width of the resonance
are $E_0=0.8$ MeV and $\Gamma(E_0)=0.6$ MeV, respectively.
The two virtual states that produce the resonance are indeed
at $|k|\sim \Mlo$.
The real bound state, for reasons that cannot be understood
from the EFT itself, turns out numerically to be 
at considerably higher momentum, 
where the EFT can no longer be trusted.
This is consistent with the known absence of a real $^5$He bound state.

We also illustrate in Fig.~\ref{fig:nhe4P32} an important
aspect of the power counting.
The dotted line shows the result from iterating
$C_2^p$ alone. In other words, it is the contribution
of the scattering length
only. This curve, which would come from a naive application of
the power counting for $s$ waves \cite{vKo99,KSW98},
does not correspond to any order in the power counting developed
here, and  clearly fails to describe the resonance near $E_n=1$ MeV.

\begin{figure}[tb]
\begin{center}
\includegraphics[width=4in,angle=0,clip=true]{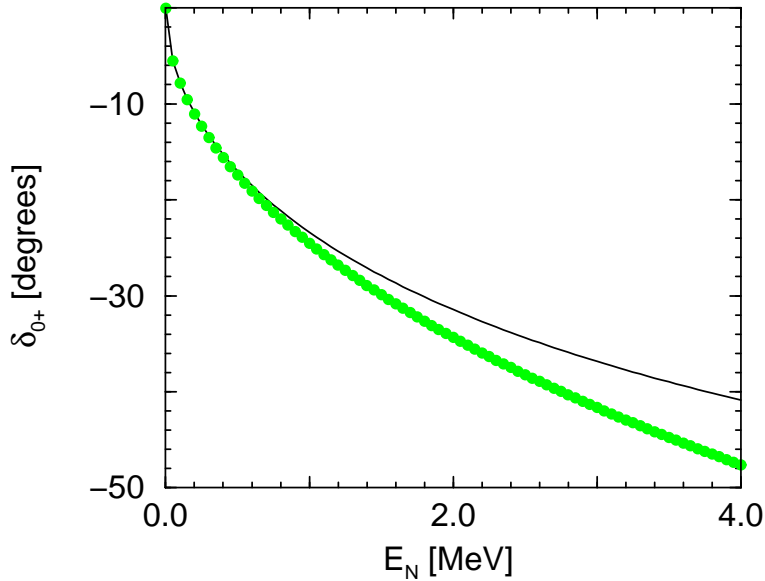}
\end{center}
\vspace*{-0pt} 
\caption{The phase shift for $n\alpha$ scattering in the 
 $0+$ partial wave as a function of the neutron kinetic energy
 in the $\alpha$ rest frame.
 The solid line shows the EFT result at NLO; at LO this phase
 shift is zero. The filled circles show the result of the  
 phase-shift analysis \protect\cite{ALR73}.
 }
\label{fig:nhe4S12}
\end{figure}

In Fig.~\ref{fig:nhe4S12}, we show the phase shifts for
elastic $n\alpha$ scattering in the $0+$ partial wave 
as a function of the neutron kinetic energy
in the $\alpha$ rest frame.
%%%%%%%%%%%%%%%%%%%%%%%%%%%%%%%%%%%%%%%%%%%%%%%%%%%%%%%%%% 
%%%%%%%%%%%%%%%%%%%%%%%%%%%%%%%%%%%%%%%%%%%%%%%%%%%%%%%%%%%%%%
In LO the phase shift is zero.
The solid line shows the EFT result at 
next-to-leading order. The NLO result already shows good
agreement with the full phase-shift analysis \cite{ALR73},
depicted by the filled circles.

The phase shifts in the $1-$ and all other partial waves are identically
zero to NLO. The first non-zero contribution appears at N$^3$LO 
in the $1-$ channel. All other waves appear at even higher orders.
That they are indeed very small one can conclude from
their absence in the phase-shift analysis \cite{ALR73}.

\begin{figure}[tb]
\begin{center}
\includegraphics[width=4in,angle=0,clip=true]{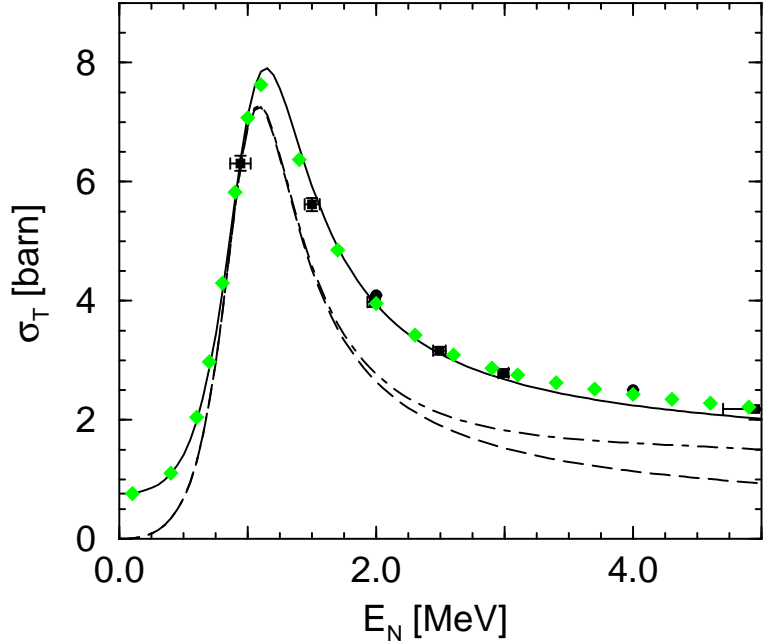}
\end{center}
\vspace*{-0pt} 
\caption{The total cross section for $n\alpha$ scattering
 in barns as a function of the neutron kinetic energy in the
 $\alpha$ rest frame. The diamonds are evaluated data from
 Ref.~\cite{BNL}, and the black squares are experimental 
 data from Ref.~\cite{data}.
 The dashed and solid lines show the EFT results at 
 LO and NLO, respectively. The dash-dotted line shows the  LO
 result in the modified power counting where the $1-$ partial 
 wave is promoted to leading order.}
\label{fig:sigtot}
\end{figure}

Obviously, not all partial waves are treated equally in
our power counting.
In order to further assess if the power counting is appropriate, 
we compare the EFT predictions 
directly to some observables. 
In Fig.~\ref{fig:sigtot}, we compare the EFT
predictions with data for the total cross section
as a function of the neutron kinetic energy in the
$\alpha$ rest frame.
%%%%%%%%%%%%%%%%%%%%%%%%%%%%%%%%%%%%%%%%%%%%%%%%%%%%%%%%%% 
%%%%%%%%%%%%%%%%%%%%%%%%%%%%%%%%%%%%%%%%%%%%%%%%%%%%%%%%%%%%%%
The diamonds are ``evaluated data points'' from
Ref.~\cite{BNL}.
In order to have an idea of the error bars from individual experiments
we also show data from Ref.~\cite{data} as the black squares.
The dashed line shows the EFT result at LO which
already gives a fair description of the resonance region but
underestimates the cross section at threshold.
The NLO result given by the solid curve gives a good
description of the cross section from threshold up to energies of 
about 4 MeV.

\begin{figure}[tb]
\begin{center}
\includegraphics[width=4in,angle=0,clip=true]{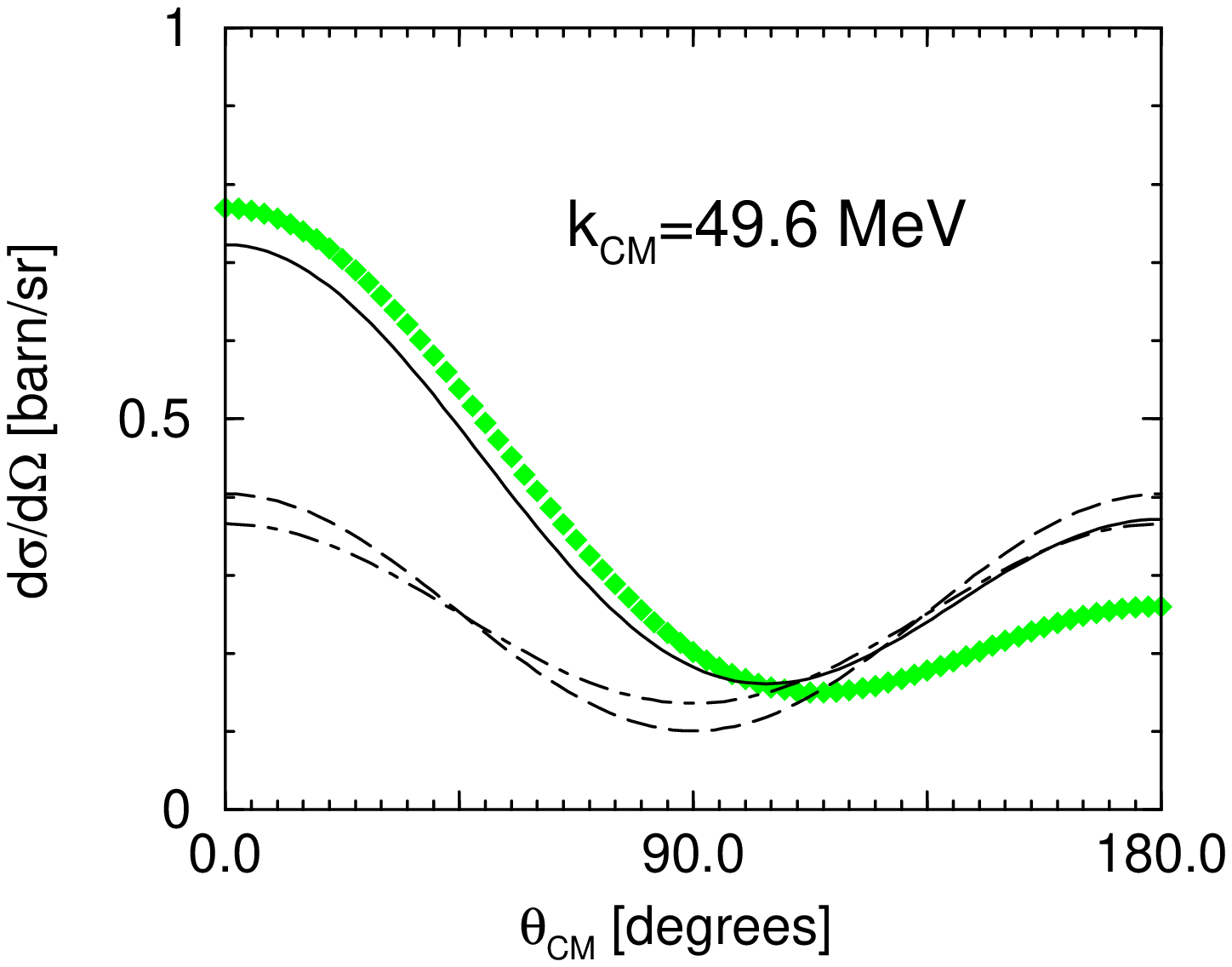}
\end{center}
\vspace*{-0pt} 
\caption{The differential cross section for $n\alpha$ 
 scattering in the center-of-mass frame
 in barns/sr as a function of the scattering angle $\theta_{CM}$
 at a momentum $k_{CM}=49.6$ MeV. 
 The diamonds are evaluated data from Ref.~\cite{BNL}.
 The dashed and solid lines show the EFT results at 
 LO and NLO, respectively. The dash-dotted line shows the  LO
 result in the modified power counting where the $1-$ partial 
 wave is promoted to leading order.}
\label{fig:diffsig}
\end{figure}

We can also calculate other observables.
As another example, we show in Fig.~\ref{fig:diffsig}
the center-of-mass differential cross section at a 
momentum $k_{CM}=49.6$ MeV. (This 
corresponds to a neutron kinetic energy $E_n=2.05$ MeV
in the $\alpha$ rest frame.)
%%%%%%%%%%%%%%%%%%%%%%%%%%%%%%%%%%%%%%%%%%%%%%%%%%%%%%%%%% 
%%%%%%%%%%%%%%%%%%%%%%%%%%%%%%%%%%%%%%%%%%%%%%%%%%%%%%%%%%%%%%
The diamonds are evaluated data from Ref.~\cite{BNL} 
\footnote{In order to obtain the differential cross section 
from the NNDC neutron emission spectra 
we divide by 2$\pi$ and multiply by the total cross section.}. 
The dashed line show the EFT results at 
LO, which is pure $p$ wave. At NLO, shown as a solid line,
interference with the $s$-wave term gives essentially the correct shape.

If we carry out the EFT to a sufficiently high order,
we will have included all terms used in the 
phase-shift analysis \cite{ALR73}, and more.
At this order, the high quality of our fit
is purely a consequence of the high quality of that fit.
Note, however, that this is by no means true at the lower 
orders explicitly displayed above.
In particular, it is perhaps surprising that our $1-$ wave does
not appear until relatively high order. 
The fact that the EFT converges fast to data
shows that the power counting developed here is reasonable.
The $1-$ wave is further discussed in the next section.

\subsection{Further Discussion of the Power Counting}
\label{further}

As we have shown, the EFT describes the data
pretty well at least up to $E_n=4$ MeV or so.
One way to improve the convergence at higher energies
is to take the scale of non-perturbative
phenomena in the $1-$ wave as a low scale. We can 
modify the power counting and count the $1-$ parameters the same as
the $1+$ parameters. The LO Lagrangian from Eq.~(\ref{lagLO}) then 
has an additional term
\bea
{\cal L}_{\rm ALO}&=&
\eta_{1-} d^\dagger\bigg[i\partial_0+\frac{\nab^{\,2}}{2(m_\alpha+m_N)}
-\Delta_{1-} \bigg] d \nonumber\\
&+& \frac{g_{1-}}{2}\bigg\{ d^\dagger \vecsigma^\dagger \cdot
\bigg[N \nab \phi-(\nab N)\phi\bigg]+ {\rm H.c.} -r\bigg[d^\dagger 
\vecsigma^\dagger \cdot\nab(N \phi)+ {\rm H.c.}\bigg]\bigg\}\,.
\label{lagALO}
\eea
The calculation of the $T$-matrix for the $1-$ partial wave proceeds 
exactly as for the $1+$ partial wave. The amplitudes $F$ and $G$
acquire the following additional contributions at leading order
\bea
F^{\rm ALO}(k,\theta)&=&
\frac{k^2\cos\theta}{-1/a_{1-}+r_{1-}k^2/2-ik^3}\,,
\nonumber\\
G^{\rm ALO}(k,\theta)&=&
-\frac{k^2\sin\theta}{-1/a_{1-}+r_{1-}k^2/2-ik^3}\,.
\eea

\begin{figure}[tb]
\begin{center}
\includegraphics[width=4in,angle=0,clip=true]{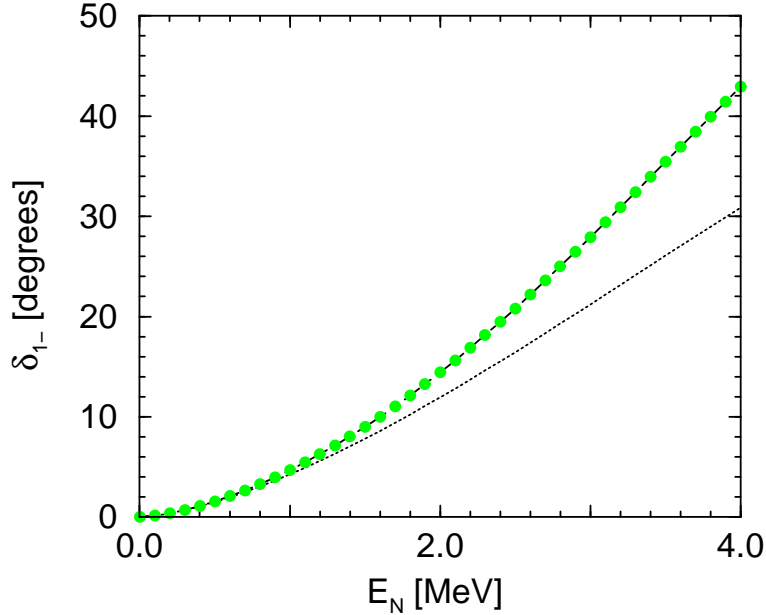}
\end{center}
\vspace*{-0pt} 
\caption{The phase shift for $n\alpha$ scattering in the 
 $1-$ partial wave as a function of the neutron kinetic energy
 in the $\alpha$ rest frame. The filled circles show the result of the 
 phase-shift analysis \protect\cite{ALR73} which is exactly reproduced
 by leading-order EFT with the modified power counting given by the
 dash-dotted line.
 The dotted line shows the contribution of the scattering length only.}
\label{fig:nhe4P12}
\end{figure}

In Fig.~\ref{fig:nhe4P12}, we show the phase shifts for
$n\alpha$ scattering in the $1-$ partial wave obtained
in this alternative power counting.
%%%%%%%%%%%%%%%%%%%%%%%%%%%%%%%%%%%%%%%%%%%%%%%%%%%%%%%%%% 
%%%%%%%%%%%%%%%%%%%%%%%%%%%%%%%%%%%%%%%%%%%%%%%%%%%%%%%%%%%%%%
The filled circles show the result of the partial-wave analysis of
Ref.~\cite{ALR73} which is exactly reproduced by the leading-order 
EFT with the modified power counting given by the dash-dotted line.
The dotted line shows the contribution of the scattering length only.
The next-to-leading order in the modified counting cannot be 
easily computed at present because ${\cal P}_{1-}$ is not known.

The cross sections corresponding to the leading order in 
the modified power counting are shown by the dash-dotted curve
in Figs.~\ref{fig:sigtot} and \ref{fig:diffsig}. 
In the total cross section, promoting the $1-$ partial 
wave to leading order gives almost no improvement compared to
the original counting except at higher energies, 
but even there the NLO result in the original counting gives 
better results. In the differential cross section at
$k_{CM}=49.6$ MeV (corresponding to
$E_n=2.05$ MeV in the $\alpha$ rest frame), the alternative
power counting gives no improvement over the LO result
(compare the dashed and dash-dotted lines in Fig.~\ref{fig:diffsig}). 
We did not find a significant improvement in the 
differential cross section over the leading order by promoting the $1-$
partial wave for neutron energies up to $E_n\approx 4$ MeV.
For reproducing the differential cross section, the interference
between $s_{1/2}$ and $p_{3/2}$ waves is much more important than the 
additional $p_{1/2}$ contribution.
As a consequence, we deem the original power counting 
most appropriate for elastic $n\alpha$ scattering
at $k\sim \Mlo$.

Finally, note that for $k\ll \Mlo$
the power counting has to discriminate between momentum
$k$ and the low-energy scale  $\Mlo$.
The $p$ waves, for example, die faster than the $s$ waves.
That is the reason our results for the 
cross section in this region are not good until we get to NLO.
It is easy to adapt the power counting for $k\ll \Mlo$:
in fact, the full amplitude
---all waves, that is--- can be treated in perturbation theory,
as in Sect. \ref{natcase}.
For more details, see Ref.~\cite{vKo99}.

\section{Conclusion and Outlook}
\label{sec:conc}

In this paper we have examined the problem of the interaction
between a neutron and an $\alpha$ particle at low energies.
We showed that a power counting can be formulated that
leads to consistent renormalization.
In leading order, two interactions have to be fully iterated.
These two interactions generate a shallow $p_{3/2}$-wave resonance
near the observed energy and width.
In subleading orders the phase shifts in all waves
can be systematically improved.
Observables calculated directly are very well reproduced.

The crucial ingredient for the applicability of the EFT
to bound states and resonances of halo type is their low characteristic
energies.
In this sense, the deuteron can be thought of as the simplest
halo nucleus whose core is a nucleon. 
$n\alpha$ scattering plays an analogous role here
as $np$ plays in the 
nucleons-only EFT.
It is clear now how to extend the EFT to more complicated
cores:
one simply introduces an appropriate field for the core under consideration,
extends the power counting to the relevant channels,
and determines the strength of interactions order-by-order
from data.

With the parameters of the nucleon-core interaction fixed in
lowest orders, we can proceed to more-body halos.
The simplest example is $^6$He. In addition to
the $n\alpha$ interaction, the $nn$ interaction has also
been determined from data.
$^6$He, like the triton, can be described as a three-body system
of a core and two neutrons.
The role of a three-body interaction can be addressed
by renormalization group techniques \cite{BHK99,ref6he}.

Note that the EFT approach is by no means restricted to neutron halos.
The Coulomb interaction can
be included in the same way as in the nucleons-only sector
\cite{em},
allowing for the analysis of nuclei such as $^8$B.
Radiative capture on halo nuclei, 
such as $p+ \mbox{$^7$Be} \to \mbox{$^8$B}+ \gamma$,
can then be calculated much like 
$n+ p\to d+ \gamma$.

Our approach is not unrelated to traditional
single-particle models.
In the latter, the nucleon-core interaction is frequently
parametrized by a simple potential with central and
spin-orbit components \cite{potrev}.
The parameters of the potential are adjusted to reproduce
whatever information is accessible experimentally.
In the EFT, we make the equivalent to a multipole
expansion of the underlying interaction.
The spin-orbit splitting, in particular,
results from the different parameters of
the dimeron fields with different spins.
In the EFT
the nucleon-nucleon interaction is 
treated in the same way as the nucleon-core interaction,
{\it mutatis mutandis}.
Contact $NN$ interactions have in fact already been used in
the study of Borromean halos \cite{george}.
It was found that density dependence, 
representing three-body effects, needed to be added
in order to reproduce results from more sophisticated parametrizations
of the $NN$ interaction. 
In the EFT, 
the need for an explicit three-body force
can be decided on the basis of the renormalization group
before experiment is confronted.
A zero-range model with purely $s$-wave $NN$ and nucleon-core 
interactions was examined in Ref. \cite{tobiaselauro}.

The EFT unifies single-particle approaches 
in a model-independent framework,
with the added power counting that allows for an {\it a priori}
estimate of errors.
It also casts halo nuclei within the same framework now used to describe
few-nucleon systems consistently with QCD \cite{Weinberg,Birareview}.
Therefore, the EFT with a core can in principle be matched
to the underlying, nucleons-only EFT.
Nuclei near the drip lines open an exciting new field 
for the application of EFT ideas.
It remains to be seen, however, whether these developments will prove
to be a significant improvement over more traditional approaches.

\section*{Acknowledgments}
We would like to thank Martin Savage for an interesting question,
and Henry Weller and Ron Tilley for help in unearthing 
$n\alpha$ scattering data.
HWH and UvK are grateful to the Kellogg Radiation Laboratory of Caltech
for its hospitality,
and to RIKEN, Brookhaven National Laboratory and to the U.S.
Department of Energy [DE-AC02-98CH10886] for providing the facilities
essential for the completion of this work.
This research was supported in part by the National Science
Foundation under Grant No.\ PHY-0098645 (HWH)
and by a DOE Outstanding Junior Investigator Award (UvK).    

\end{document}